# Applied User Research in Virtual Reality: Tools, Methods, and Challenges

Leonie Bensch*, Andrea Casini*, Aidan Cowley*, Florian Dufresne*, Enrico Guerra*, Paul de Medeiros*, Tommy Nilsson*, Flavie Rometsch*, Andreas Treuer*, and Anna Vock*

*All authors have contributed equally to this work

**Abstract**
This chapter explores the practice of conducting user research studies and design assessments in virtual reality (VR). An overview of key VR hardware and software tools is provided, including game engines, such as Unity and Unreal Engine. Qualitative and quantitative research methods, along with their various synergies with VR, are likewise discussed, and some of the challenges associated with VR, such as limited sensory stimulation, are reflected upon. VR is proving particularly useful in the context of space systems development, where its utilisation offers a cost-effective and secure method for simulating extraterrestrial environments, allowing for rapid prototyping and evaluation of innovative concepts under representative operational conditions. To illustrate this, we present a case study detailing the application of VR to aid aerospace engineers testing their ideas with end-users and stakeholders during early design stages of the European Space Agency's (ESA) prospective Argonaut lunar lander. This case study demonstrates the effectiveness of VR simulations in gathering important feedback concerning the operability of the Argonaut lander in poor lighting conditions as well as surfacing relevant ergonomics considerations and constraints. The chapter concludes by discussing the strengths and weaknesses associated with VR-based user studies and proposes future research directions, emphasising the necessity for novel VR interfaces to overcome existing technical limitations.

**Introduction**
      Interactive simulations in Virtual Reality (VR) provide the means for rapid assessment and iterative development of novel design concepts without incurring many of the financial, logistical or safety constraints typically associated with real-world studies of working prototypes. This has made VR into a particularly useful tool in the field of space systems design, due to its capacity to simulate relevant extraterrestrial environmental conditions that cannot be replicated through other means in the real world, whilst also helping to reduce some of the massive costs normally entailed in development of novel space systems.
      Such VR simulations enable aerospace engineers to visualize and continuously test relevant ideas with a wide range of end-users and expert stakeholders from the earliest stages of a design process. The reflections, behavioural measurements and observations acquired through such user studies can subsequently help steer concept optimization and further decision-making, whilst also allowing engineers and designers to gain a better understanding of the underlying problems in order to adjust or redefine their goals as needed. This, in turn, allows for rapid iteration of solutions during early

design stages, thereby reducing the risk of having to implement expensive changes further down the development process.

Following this line of reasoning, the European Space Agency (ESA) has been drawing on VR to help guide the ongoing development of the prospective Argonaut lunar lander. In this chapter we offer a first-hand account of this endeavour, detailing ESA's approach and its outcomes. By reflecting on the technical limitations inherent in conventional VR interfaces, such as insufficient haptic feedback, we elaborate not just the opportunities, but also the challenges involved in using VR to conduct valid user studies. In doing so, we provide lessons learned from ESA's work that may help future practitioners responsibly harness some of the considerable potential offered by VR in space engineering and beyond.

**Tools**

VR technology has evolved significantly over the past few decades, becoming more widely available and accessible. Early advancements date back to the 1960s when Ivan Sutherland developed one of the forerunners of modern VR headsets, featuring head-tracking and the display of computer-generated images (Sutherland, 1968). NASA's Virtual Interface Environment Workstation (VIEW) introduced in 1988 further pushed the boundaries of VR with its advanced capabilities for immersive and highly interactive simulations (Fisher et al., 1988). The potential offered by this technology quickly became apparent when it played a pivotal role in training the Hubble space telescope flight team for a critical repair mission in 1993 (Loftin & Kenney, 1995).

Aside from head-mounted displays (HMDs), another noteworthy VR approach is the CAVE (Cave Automatic Virtual Environment) technology. CAVEs surround the user with screens projecting virtual images onto walls and the floor, eliminating the need for discomforting headsets. Additionally, participants in multi-user applications can physically see each other, enhancing social interaction (Manjrekar et al., 2014). However, due to the complexity and cost associated with setting up CAVE systems, their widespread adoption has been limited. This has resulted in the prevalence of modern HMDs in most domains, including research, industry, and the private sector.

Consumer-grade VR headsets, such as the Meta Quest 2 (Meta, 2020), HTC Vive (HTC, 2023), and Pico 4 (Germany, 2023) have gained popularity due to their affordability, high visual fidelity (up to 2448 x 2448 pixels per eye (Deutschland, 2023)), and ease of use. Standalone headsets, exemplified by the Meta Quest 2, don't require an external PC or tracking devices due to their inside-out tracking capabilities. They can be easily used by simply powering them on and placing them on the head (connection to a PC for performance-consuming applications is still possible with most models). On the other hand, PC-supported headsets like the HTC Vive Pro ( utilise lighthouse tracking, an infrared-based technology developed by Valve. These headsets require installation of external tracking devices before use.

Interacting with VR applications commonly involves the use of controllers, which resemble game console controllers with various input keys assigned to specific functions and actions. Alternative input methods, such as haptic gloves (Perret & Vander Poorten, 2018) or hand tracking and gesture control (Buckingham, 2021), also exist. However, the controller-based interaction scheme remains prevalent in the majority of VR applications. While VR hardware is crucial, the significance of software and virtual experiences cannot be overstated. Game engines such as Unity (Unity, 2023) and Unreal Engine (Engine, 2023) play a vital role in creating applications for modern VR headsets. These engines offer comprehensive development environments that enable the creation of complex three-dimensional simulations and immersive experiences. Unity and Unreal Engine, in particular, stand out due to their extensive VR functionalities, support for common VR platforms, and large developer communities. These engines offer comprehensive toolsets that simplify otherwise complex technical processes like performance optimization, interaction design, and realistic physics simulation. Additionally, their asset import pipelines facilitate seamless integration with various digital creation tools, including 3D modelling and texturing software. As a result, they enable developers to quickly and effectively model hypothetical design solutions and relevant scenarios, which can subsequently be explored by relevant stakeholders and assessed through user research methods.

**Methods**

Much like real-world studies, user research conducted in VR can be broadly categorised into qualitative and quantitative approaches. Qualitative approaches primarily strive to generate a comprehensive in-depth understanding and interpretation of users' needs, actions, behaviours, beliefs, and emotions, unearthing the processes and meanings that drive them. As such, they offer insights into the "why" of user behaviours or beliefs (Hennink et al., 2020. This is typically achieved through the application of descriptive research methods, including, but not limited to, interviews and focus groups. Such studies, particularly in the highly specialised context of space systems design, typically involve a relatively small, handpicked sample of participants, such as domain experts. By leveraging the informed perspectives of these participants, researchers gain a unique opportunity to develop understanding of relevant design issues from the viewpoint of potential users and other key stakeholders. The immersive nature of VR enabling participants to engage in hypothetical situations while being put in the shoes of prospective users, (e.g., future astronauts), can serve as a fertiliser for such qualitative enquiries. Indeed, its capacity to foster empathy and provide an in-depth understanding of diverse user perspectives has led to VR being dubbed the "ultimate empathy machine" (Herrera et al., 2018).

Similarly, the ability to convey rich immersive settings allows researchers to explore the contextual factors influencing user's behaviours and perspectives. This is particularly useful in projects concerned with environments that are difficult to access in the real world, such as space systems design, where frequent evaluations of concepts in remote space environments (e.g., the Moon, Mars, International Space Station [ISS]) are necessary. In doing so, VR effectively endows design assessments with elements of contextual inquiry (Duda et al., 2020), putting the spotlight on synergies and frictions arising between the depicted design solutions, users and environments (Nilsson et al., 2023). Methodologies employed to elicit such reflections typically pivot around reflections gathered through the think-aloud protocol, as well as interviews and collaborative design sessions conducted with the user immersed in the VR environment.

In contrast to qualitative research, quantitative research aims to interpret phenomena through statistical patterns. This approach involves a larger sample size and seeks to derive findings that can be generalised to a broader population. By collecting measurable numerical data and employing statistical techniques, quantitative methods equip researchers with the means to address questions such as "how much", "what," and "where" (Apuke, 2017 ).

The requisite numerical data is primarily collected through experiments or user surveys, which are then scrutinised using statistical methods, including correlation or regression analysis. This process enables the clear affirmation or rejection of pre-formulated hypotheses based on the statistical findings. In the realm of space system design, quantitative research in VR is particularly useful for conducting comparative studies of multiple potential solutions (A/B testing) or evaluating users' physiological responses.

Common techniques employed in this domain include task performance measurements (such as completion times and error rates), analysing self-reported survey responses (for example, using Likert scales to measure workload via the NASA TLX scale (Hart, 2006)), eye-tracking (for attention, information processing, user engagement), motion capturing (mocap), and biometric measurements (e.g. heart rate or brain activity) (Becker et al., 2022).

However, it's worth noting that qualitative and quantitative methods are not mutually exclusive. The most comprehensive insights often emerge from a hybrid, or *mixed-methods*, approach (Sandelowski, 2000). For example, qualitative user reflections might be gathered through a think-aloud protocol and semi-structured interviews during immersive VR sessions in order to better understand the domain experts' perspectives. Simultaneously, quantitative data may be collected to gain insights into their physiological responses (e.g., heart rate) to various stimuli present in the VR environment.

**Challenges**

Despite the numerous valuable applications of VR simulations, several challenges persist that hinder the utilisation of VR to its full potential in the field of space systems design. While VR technology has made significant advancements in visual fidelity, there is a prevalent concern about its limited ability to stimulate other sensory modalities, which may undermine the validity of observations made in VR environments. Major discrepancies between human responses in VR environments and their real-world equivalents have been linked to such technical limitations. Notably, the so-called "super soldier syndrome" causes users to exhibit game-like attitudes, such as careless behaviour and unrealistic risk-taking, thus potentially invalidating behavioural user data collected during such studies (Barlow & Morrison, 2005b).

Technical difficulties associated with providing accurate haptic feedback, in particular, have been shown to impede perceived realism of VR environments, thus posing a considerable challenge for user research activities. Given the important role of altered sensory experiences in space, such as the sensation of hypogravity, movement restrictions incurred by bulky space suits, inertia, and other haptic or tactile sensations, the space systems design domain may be particularly affected by such limitations, potentially restricting the applicability of VR-based research approaches in ongoing design and development endeavours.

Instead, relevant conditions, such as hypogravity, have traditionally been simulated using real-world neutral buoyancy facilities or testbeds on the seafloor (Koutnik et al., 2021), as well as parabolic flight campaigns (De Martino et al., 2020). Such practices come with their own limitations: neutral buoyancy facilities require equipment to be specifically modified for the right buoyancy in order to simulate appropriate gravity levels, and the hypogravity conditions experienced during parabolic flights typically only last for approximately 20-30 seconds. Moreover, such approaches likewise entail oftentimes prohibitively high costs.

There is a need, then, for novel and efficient interfaces capable of compensating for some of the limitations associated with VR. One technology in this vein that has shown promise is a gravity offload system. NASA's ARGOS, for instance, has been proven capable of recreating the effects of desired gravity levels by offloading VR users by a desirable factor, thereby more reliably simulating environmental conditions on the Moon, Mars or the ISS (Orr et al., 2022).

Similarly, whilst spacesuits can be visually simulated in a VR environment, the movement restrictions associated with an actual suit cannot be replicated digitally. Given the limited access to actual spacesuits due to their considerable weight and cost, it is vital for future studies to identify adaptable and efficient simulation methods. Potential solutions include the development of cheaper spacesuit mock-ups that realistically recreate some of the relevant movement constraints. Here, systems such as the Austrian Space Forum Aouda.X space suit simulator (Groemer et al., 2012 ) or the Extravehicular Activity Space Suit Simulator (EVA S3) developed at the Massachusetts Institute of Technology (Meyen, 2013) could prove useful to provide realistic and cost-effective means to simulate movement restrictions in virtual environments.

To simulate haptic sensations, haptic gloves (Zhu et al., 2020) or suits can be used to replicate forces and tactile sensations. Another approach involves combining physical mock-ups with virtual visual content, as demonstrated in NASA's hybrid reality lab (Delgado & Noyes, 2017).

**The Argonaut Case Study**

The notion of reviving crewed missions to the Moon and establishing a permanent human presence on its surface is rapidly gaining prominence among private entities and governmental space agencies alike. The fruition of this ambition will hinge on the development and implementation of robust logistics systems capable of ensuring a dependable delivery of vital supplies and cargo to the lunar surface in support of future human expeditions.

Against this backdrop, the ESA is currently engaged in a feasibility study aimed at laying the foundation for future development of the Argonaut lunar lander. The primary objective of this autonomous lunar landing vehicle will be the transportation of diverse crew supply payloads and scientific experiments to the lunar surface. With an anticipated initial deployment in 2029, Argonaut is poised to play a key role in Europe's pursuit of sustainable human exploration and colonisation of the Moon.

Primary aims of the early stages of this project include elaborating known design challenges, uncovering novel ones, and validating potential design solutions. Given the nascent nature of this endeavour, with no physical prototype of the Argonaut having been constructed so far, VR has been deemed a fitting tool for fulfilling these objectives.

Consequently, our team was tasked to construct a virtual prototype of the Argonaut lander which could then be subjected to assessments using interactive VR. Drawing on input from the Argonaut project management and responsible engineering teams, we produced a representative configuration of the lander (Nilsson et al., 2022).

The result was an approximately 2.8 metres tall virtual lander featuring an octagonal cargo deck of approximately 14 square metres on top. In order to foster comprehensive discussions encompassing the broader operational context and to facilitate the examination and evaluation of aspects concerning usability and human factors challenges and limitations, we strived to incorporate an additional level of detail into the model, such as communication antennas and radiators, as well as a set of 4 cargo containers located at the cargo deck. A ladder and transportation cart were likewise included. The VR users were also embodied in a 3D model of the Exploration Extravehicular Mobility Unit (xEMU) EVA suit. For a schematic drawing of the lander mock-up and accompanying hardware, see Figure 11.1.

In addition to creating the virtual Argonaut mock-up and relevant hardware, we also needed to construct a representative lunar environment wherein the lander's operations could be simulated and evaluated. To accomplish this, we utilised topographic scans of the lunar surface captured by NASA's Lunar Reconnaissance Orbiter (LRO) (Smith, D. E. et al., 2017). Drawing on these scans, we reconstructed digitally an 8 by 8 kilometres area in close proximity to the Shackleton crater situated at the Lunar south pole (89.9°S 0.0°E). We selected this particular location due to it having been identified as one of the potential landing sites for the initial Artemis human landing ( Smith et al., 2020), thereby providing a credible setting for our Argonaut assessment.

The LRO topographic scans had a relatively low resolution, with each pixel representing an area of 100 by 100 metres on the real lunar surface. Consequently, finer surface details (e.g., boulders or smaller craters) had to be recreated manually using conventional 3D modelling software, such as Cinema 4D. Additionally, we developed a custom terrain shader to enhance the visual realism of the virtual lunar surface. Throughout this process, we collaborated closely with an experienced lunar geologist to ensure the utmost accuracy and authenticity of our virtual lunar landscape.

Accurate replication of the unique lighting conditions on the lunar south pole was another crucial aspect of our work. The virtual sun (i.e., a directional light in the VR environment) was oriented towards the north, set at an angle of 1.5° above the horizon, and its intensity was adjusted to 1.37 kW/m$^2$ (Vanoutryve et al., 2010). To faithfully reproduce the absence of a lunar atmosphere, all forms of indirect lighting and light scattering were deactivated, resulting in the creation of deep, dark shadows and blinding highlights that typify the lunar environment.

Once the development of the virtual lunar landscape was completed, we positioned the Argonaut mock-up at its centre. Finally, to bring the entire VR experience to life, we utilised the Unreal Engine 4 game engine. By combining the realistic lunar environment, the detailed Argonaut mock-up, and the immersive capabilities of the game engine, we thus provided an authentic and engaging VR simulation for evaluation and assessment purposes.

To facilitate such an evaluation, we hand-picked and invited a number of experts from the field of human spaceflight to experience our VR simulation of the Argonaut mock-up. Notably, our study included two active astronauts, both of whom have extensive experience conducting EVA operations outside the International Space Station (ISS). Other participants included engineers responsible for providing technical support for ISS, including the maintenance of several of its modules. Multiple participants had likewise experience with monitoring of astronaut missions through the European

Spacecraft Communicator (EUROCOM) - an international flight control centre responsible for direct communication with the ISS crew.

Each participant experienced the VR simulation individually. Upon being briefed about the study purpose, the participant was asked to complete a questionnaire regarding their demographics and previous experience with VR. After a quick demonstration of the VR controllers, the participant was then given freedom to navigate the virtual environment and explore the Argonaut lander at leisure. Following the think-aloud protocol (Ericsson & Simon, 1980), participants were encouraged to verbalise their thoughts and reasoning while completing this examination. Once done examining the lander, participants were asked to answer a set of semi-structured interview questions. These questions were open-ended and aimed to elicit participants' reflections and comments on their experience. Specific topics of inquiry included various features of the lunar lander, such as the cargo unloading mechanism, antenna placement, design of the cargo containers, ladder, and transportation cart. Participants were also asked to identify potential safety hazards and suggest design improvements. The study did not impose any time limits, allowing participants to take as much time as needed to explore the lander and respond to our questions. As a result, the total duration of the sessions varied between 40 to 80 minutes.

Participant responses were documented through notes, audio recordings, and questionnaire responses. The dataset was then independently coded by three researchers, and any inconsistencies were resolved through discussion. The data were synthesised using a qualitative thematic analysis (Braun & Clarke, 2006).

A recurring theme throughout the study were concerns raised by our participants about the lighting conditions on the Moon and its impact on various design aspects of the Argonaut lander. The Argonaut's landing legs, for instance, were seen as potential tripping hazards in poor lighting. Similarly, locating misplaced payload containers and equipment was likewise seen as problematic in shadowed areas. Suggestions included equipping mobile hardware with artificial light sources and making greater use of LED bars or light strips.

Conversely, in other areas, the blinding sunlight, exacerbated by the absence of the Moon's atmosphere, posed additional challenges. Participants expressed concerns, for instance, about being blinded and experiencing momentary vision impairment while performing EVA procedures. Reflections from the Argonaut's aluminium and metallic components were also problematic, with anti-glare coatings and sun visors proposed as potential remedies.

Solar power generation and the ideal placement of solar panels on the lander also frequently sparked discussion. Participants considered scenarios where the lander inadvertently lands in a shadow, rendering solar power generation impossible, and proposed countermeasures, such as mounting solar panels on tethered transportation carts for repositioning. Matters concerning temperature management and solar power generation in the design of future landers and other lunar infrastructure were likewise frequently brought up.

Another frequent topic of reflection were the dimensions and ergonomic suitability of design elements, particularly the cargo containers. Our study participants, for example, found that the container handles were too slim to allow efficient manipulation during extravehicular activities, given the bulky gloves that astronauts would likely be wearing. Limited visibility while manually carrying the large containers and potential difficulties in maintaining balance were also identified as key design considerations. Redesigning container handles by, for instance, adding attachment mechanisms or wheels, and using rectangular containers (to improve tileability) were suggested as possible solutions.

The notion of astronauts having to climb up and operate on top of the Argonaut cargo deck likewise attracted numerous comments pertaining to potential safety and visibility issues. Recommendations included adjusting ladder angles, using side-view mirrors, and implementing protective measures for sensitive instruments.

As expected, the study also surfaced technical limitations of VR that may have influenced some of the participants' comments. The lack of haptic feedback hindered accurate assessment of weight and object manipulation. The absence of movement constraints imposed by bulky spacesuits also affected the evaluations. When prompted to elaborate such limitations, our participants generally argued the VR simulation was more suitable for providing situational awareness rather than for assessing physical mechanics. In particular, participants appreciated the contextual experience and found VR useful for planning EVA operations and visualising robotic actions. Despite its limitations, VR was thus recognized as a valuable design tool with strengths to aid future design activities.

**Takeaways and Future Directions**
As demonstrated in the outlined case study, VR can be purposefully applied to facilitate user studies during early design stages. Two of its qualities make it particularly suitable for Space Systems Design - the capacity to simulate any chosen environment, including extreme environments, in a visually compelling and highly accessible manner, as well as its suitability to integration with other digital systems. More specifically, intelligibly conveying adverse conditions in a cost-effective, flexible, and, above all, safe way can provide invaluable insights into early-stage designs and operational concepts. Such adverse conditions may include disruptive lighting, rough terrain, scale- and distance-gauging difficulties, as well as low gravity conditions and their associated consequences, to give some examples.

VR's suitability for integration with broader operational ecosystems, on the other hand, enables model-based systems engineering project approaches, allowing for correct evaluation of, and interaction with, digital twins of spaceflight systems. Owing to VR's accessible digital interfaceability, real-time integration of live data such as satellite observations, spacecraft telemetry or other forms of state information from ongoing missions becomes significantly more practical for such applications.

While effectively allowing for varied visual inputs, the lack of multisensory - particularly physical - interactions remains a problem. Although an astronaut on the lunar surface may be able to lift objects weighing upwards of a hundred kilograms, the non-trivial dynamics involved in this situation cannot be intuitively conveyed through purely visual cues as offered by conventional VR interfaces. Yet, understanding effects resulting from the object's inertia, the astronaut's own reduced weight as well as motion constraints imposed by an EVA suit are essential for planning an operational concept including such activities. Similarly, pure VR studies have limitations in terms of evaluating the ergonomics of spaceflight systems beyond non-contact aspects, such as accessibility or lighting assessments. In the future, approaches combining VR simulations with additional physical interfaces, might serve to close that gap. For this purpose, we are investigating the use of various haptic devices, as well as other technology capable of introducing physical interactions into VR. As an example, ESA and the German Aerospace Center (DLR) are currently developing and building a novel lunar analogue facility named LUNA in Cologne, Germany (Casini et al., 2023). This facility will allow us to replicate some of the crucial harsh environmental conditions on the Moon such as reduced gravity conditions via a gravity offloading system, a dusty surface using a regolith simulant and the challenging lighting conditions using a Sun simulator. LUNA is meant to serve as a training and technology centre where hardware, protocols and concepts of operations can be tested in a realistic lunar setup. Furthermore, extended reality (XR) simulations are also envisioned to further enhance the immersion levels and to advance VR-based technologies for design and simulations studies. For smaller-scale and more flexible ad-hoc campaigns, solutions employing stand-alone haptic devices, such as exoskeletal gloves, tactile feedback systems, as well as effective use of spatially tracked props (weight simulators, generic contact surfaces and physical interfaces) are currently under investigation. These can be applied in a more general and varied way to VR scenarios, and their integration adapted on the fly. Additionally, cutting-edge manufacturing processes and technologies, such as additive manufacturing and laser cutting, can be leveraged to achieve prop interface specificity as required, without incurring substantial financial and labour costs. Moreover, they allow for fast iteration on a level closely approaching that of VR.

**Conclusions**

In this chapter, we have provided a broad overview of VR-based user research practices, with a particular focus on the space systems design domain. The presented Argonaut study constitutes one of the first cases of VR technology being used to ameliorate the overall design process of a future lunar system. Reflecting on the study outcomes, we have explored not just the efficacy of VR to elicit relevant feedback, but likewise some of the limitations stemming from the predominantly audiovisual nature of conventional VR interfaces. Overcoming these challenges and responsibly integrating VR into future design and engineering processes could have a transformative impact. Beyond simply reducing some of the associated costs, the highly accessible nature of VR makes it well positioned to help make space systems design activities accessible to professionals who may have previously been excluded from such projects. By enabling a wider range of individuals to contribute to the generation of ideas, designs, and solutions that will shape our future beyond Earth, VR thus exemplifies a disruptive technology that has the potential to foster a more inclusive, innovative, and dynamic space industry.

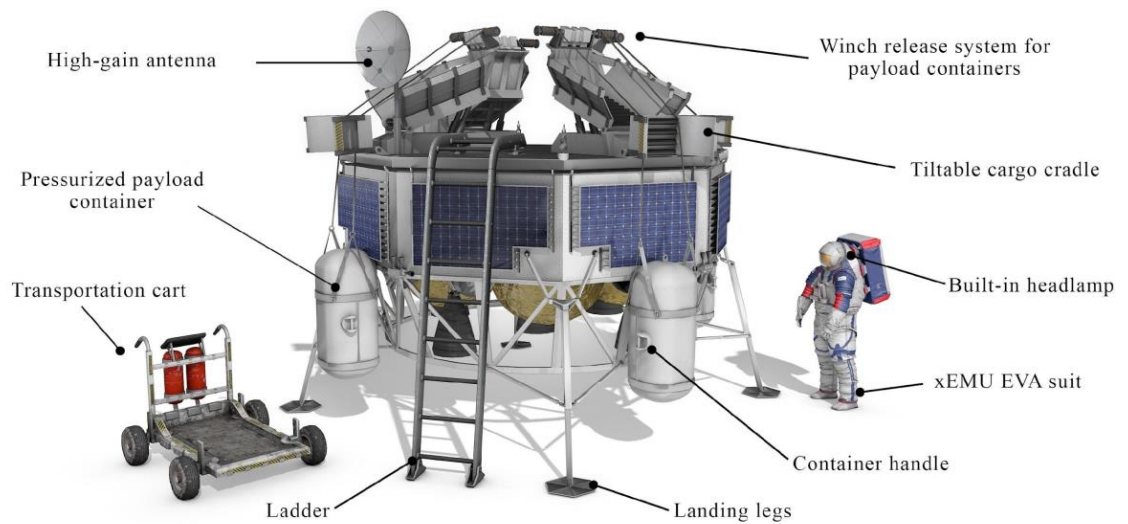

**Figure 11.1:**
*The virtual Argonaut lander mockup. Illustration adapted from (Nilsson et al., 2023).*